\documentstyle[graphicx]{mn2e}

\newif\ifAMStwofonts
\title
[Are the 2dFGRS superstructures a problem for hierarchical models?]
{Are the superstructures in the two-degree field galaxy redshift survey 
a problem for hierarchical models?}
\author [Y. Yaryura et al.]
{C. Yamila Yaryura,$^{1,2}$\thanks{Email: yaryura@mail.oac.uncor.edu} 
C. M. Baugh$^3$.
R. E. Angulo,$^4$ \\
$^1$ IATE, Instituto de Astronom{\'i}a Te\'orica y Experimental,
Laprida 851, C\'ordoba, Argentina.\\
$^2$ CONICET, Consejo Nacional de Investigaciones Cient{\'i}ficas y T\'ecnicas,
Rivadavia 1917, Buenos Aires, Argentina.\\
$^3$ Institute for Computational Cosmology, Department of Physics, University of Durham, South Road, Durham, DH1 3LE,UK.\\
$^4$ Max Planck Institut fuer Astrophysik, D-85741 Garching, Germany}
\begin{document}
\maketitle
\begin{abstract}
We introduce an objective method to assess the probability of 
finding extreme events in the distribution of cold dark matter such as voids, overdensities 
or very high mass haloes. Our approach uses an ensemble of N-body simulations 
of the hierarchical clustering of dark matter to find extreme structures. 
The frequency of extreme events, in our case the cell or smoothing volume 
with the highest count of cluster-mass dark matter haloes, is well described 
by a Gumbel distribution. This distribution can then be used to forecast the 
probability of finding even more extreme events, which would otherwise require 
a much larger ensemble of simulations to quantify. We use our technique to 
assess the chance of finding concentrations of massive clusters or superclusters, 
like the two found in the two-degree field galaxy redshift survey (2dFGRS), using a counts-in-cells 
analysis. The Gumbel distribution gives an excellent description of the 
distribution of extreme cell counts across two large ensembles of simulations covering 
different cosmologies, and also when measuring the clustering in both real 
and redshift space. We find examples of structures like those found in the 2dFGRS in the simulations. 
The chance of finding such structures in a volume equal to that of 
the 2dFGRS is around 2 \%.
\end{abstract}

\section{Introduction} 

The discovery of extreme objects, such as large voids or highly overdense 
regions called superclusters, in which many galaxy clusters are found close together, 
has often been presented as a challenge to the hierarchical structure formation paradigm. 
However, the main drawback of using the presence of unusual structures to rule out 
a particular model is that it is not always clear how to assess the probability 
of finding such objects. 
In this paper we introduce a new methodology to address this problem in which we 
use N-body simulations and extreme value theory to provide a quantitative assessment 
of the probability of finding rare structures in a given cosmology. 

Claims of rare structures are common in the literature. Frith et~al. (2003) argued 
that the shape of the local galaxy counts implies a underdense volume in 
the southern sky $\sim 300h^{-1}$Mpc across (see also Busswell et~al. 2004; Frith et~al. 
2005, 2006). Cruz et~al. (2005) found a cold spot in the cosmic microwave background 
radiation that is much bigger than expected in a Gaussian distribution.  
Rudnick, Brown \& Williams (2007) suggested that 
this cold spot is a secondary anisotropy, coinciding with the angular position 
of a void in a survey of radio galaxies. Swinbank et~al. (2007) found a large association of 
galaxy clusters, a supercluster of galaxies, at $z\sim 0.9$ 
in the UK Infrared Deep Sky Survey. 
Sylos Labini, Vasilyev \& Baryshev (2009a,b) argue that large scale 
density fluctuations are present in the local galaxy surveys which 
cannot be explained in current structure formation models.
There are two common problems with the 
interpretation of such results. Firstly, what is the selection function of these objects, 
which would allow us to fix the frequency of finding such structures? And, secondly, 
what exactly are we looking at? For example, in the case of an overdensity of galaxies have 
we seen one massive cluster or are we looking at a projection of smaller structures along the line of sight? 
How should we compare the observations to theoretical predictions? 

In this paper we assess how common the superclusters found in the two-degree field 
galaxy redshift survey (2dFGRS; Colless et~al. 2001, 2003) are in the CDM cosmology. 
These structures were identified as ``hotspots'' in the distribution of galaxy 
counts-in-cells (Baugh et~al. 2004; Croton et~al. 2004). 
One structure is in the NGP part of the 2dFGRS at a redshift of $z = 0.08$ 
and a right ascension of 13 hours, and the other is in the SGP region at 
$z = 0.11$ at a right ascension of 0.5 hours. The higher order moments 
of the counts are strongly influenced by the presence of these structures (Croton et~al. 2004; 
Nichol et~al. 2006). 
A subsequent analysis of galaxy groups in the 2dFGRS revealed 
that these regions contain a surprisingly large fraction of all the massive clusters 
in the survey (Eke et~al. 2004a). Of the 94 groups in the full flux limited 2dFGRS 
out to $z \sim 0.15$ with 9 or more members and estimated masses above 
$5 \times 10^{14} h^{-1} M_{\odot}$, 20 percent reside in these superclusters 
(Padilla et~al. 2004). The supercluster in the NGP region of the 2dFGRS is part of the ``Sloan Great Wall'' 
(Gott et~al. 2005). 

The 2dFGRS superclusters are by no means the largest superclusters in the local universe 
(for a list of superclusters, see Einasto et~al. 2001). The Shapley supercluster, for example, 
contains more Abell clusters than either of the 2dFGRS structures (Raychaudhury et~al. 1991; 
Proust et~al. 2006; Munoz \& Loeb 2008). However, not all of the clusters contained 
within Shapley and similar mass concentrations have measured redshifts. Many of the member 
clusters have been identified in projection, and so their actual size is open to debate 
(Sutherland \& Efstathiou 1991).  The advantage of focusing on the 2dFGRS structures is 
that they have been identified from an unbiased redshift survey which was designed to 
map a particular volume of the Universe, rather than by targetting known structures. The 
volume of space in which the superclusters are found is therefore well defined. 
Furthermore, through the construction of the 2dFGRS Percolation Inferred 
Galaxy Group (2PIGG) catalogue (Eke et~al. 2004a), there is a clear, 
objective way to connect the observed properties of the galaxy groups which 
make up the superstructures to dark matter haloes in N-body simulations. 

In this paper we use extreme value theory to assess the probability of finding 
structures like the 2dFGRS superclusters in the CDM cosmology. Previous attempts 
to address the probability of finding such structures have used small numbers of 
simulations, and so have not been able to make definitive statements. 
For example, Croton et~al. (2004) measured the moments of the galaxy cell count 
distribution in the 22 mock catalogues whose construction was described by 
Norberg et~al. (2002). None of these mocks displayed higher order moments 
that looked like those measured in the 2dFGRS, giving a probability of less than 
5\% that such a structure could arise in a CDM model. One possible way around this 
problem is to generate estimates of the error on a measurement from the data itself 
(see Norberg et~al. 2009). Around 50 such estimates are required to get an accurate 
estimate of the variance on a measurement in the case of Gaussian statistics, and this 
method is clearly not applicable to a structure which appears once or twice in the dataset. 
The method we describe in this paper is calibrated against N-body simulations and can 
be extrapolated to very low probabilities, without requiring any assumption about the 
detailed form of the underlying distribution, just its asymptotic behaviour.

The layout of this paper is as follows. In Section 2 we first give a very brief 
overview of extreme value theory, before discussing how we connect dark matter haloes 
from an N-body simulation to galaxy groups in the 2PIGG catalogue. Finally in Section 2 we 
describe the ensembles of N-body simulations that we use to measure the counts-in-cells 
distribution of massive haloes. The results of the paper are presented in Section 3, 
in which we show the impact of mass errors on the distribution of halo counts-in-cells 
and demonstrate how well extreme value theory describes the probability of finding a 
``hot'' cell. Our conclusions are given in Section 4.

\section{Method}
In this section we first give a brief introduction to extreme value 
theory (\S~2.1), before describing how we will look for superclusters like 
those in the 2dFGRS in N-body simulations (\S~2.2). The simulations we use are 
outlined in \S~2.3.

\subsection{Extreme value theory} 

Extreme value theory is used extensively by weather forecasters, seismologists 
and actuaries, but less so by cosmologists. However, there are some recent 
examples of applications to features in the cosmic microwave background 
(Mikelsons, Silk \& Zuntz 2009) and the fluctuations in the density around 
galaxies (Antal et~al. 2009). 
The main result of interest for this paper is the probability distribution 
of extreme events. In our application the extreme event is a high count of 
galaxy clusters within a volume of space covered by a cell. 
There is no need to specify the form of the underlying probability 
distribution of cluster counts. The only requirement for the extreme value  
distribution to be applicable to the extrema of this process is that 
the underlying distribution is well behaved, which means that it is 
continuous and the cumulative distribution has an inverse. There are three 
types of extreme value distribution, which are distinguished by the shape of 
the tail of the underlying distribution. The Gumbel distribution is a type 
one extreme value distribution, in which case the ``shape parameter'' of the 
distribution tends to zero (e.g. as is the case for a Gaussian or exponential 
distribution).   

Returning to the application of extreme value theory in this paper, imagine 
a population of galaxy redshift surveys. For each survey, the distribution of the 
counts in cells of galaxy clusters is measured for a particular cell size. The 
maximum count obtained within a cell is recorded for each survey. The cumulative 
distribution of the maximum count across the population of surveys is given by 
the Gumbel distribution: 

\begin{equation}
F(x;\mu,\beta) = \exp^{-\exp^{(\mu-x)/\beta}}, 
\end{equation}
where the mean of the distribution is $\mu + \gamma \beta$, 
where $\gamma = 0.577216$ is the Euler-Mascheroni constant. 
The standard deviation is given by $\beta \pi / \sqrt{6}$. 
The value of $\beta$, a parameter in the Gumbel distribution, 
is obtained from the standard deviation of maximum cell counts values 
across different realizations of the density field. Likewise, the 
other parameter $\mu$ is derived from the mean of the maximum cell counts.

\subsection{The extreme value data: 2dFGRS superclusters} 

As mentioned in the previous section, the extreme value data we 
will assess are the two ``hot'' cells in the 2dFGRS identified by 
Baugh et~al. (2004) and Croton et~al. (2004). These hot cells were 
initially identified in a volume limited sample of $L_*$ galaxies. 
On leaving out the galaxies within the two hot cells, the higher order 
moments of the count distribution had the expected form on large 
scales. The cell radius used by Croton et~al. was equivalent to a 
cube of side $40.6\,h^{-1}\,$Mpc. On cross-matching the hot cells in 
the galaxy distribution with the 2PIGG catalogue, there are $10$ groups 
in each cell with an estimated mass in excess of $5 \times 10^{14} h^{-1} M_{\odot}$ 
(Padilla et~al. 2004). 

Rather than construct mock galaxy catalogues to compare with the 
2dFGRS, we shall consider instead the counts of dark matter haloes. 
This removes a layer of theoretical uncertainty, as there is no longer a 
need to include a model of galaxy formation, which may in any case have 
produced a group catalogue with different properties from those of the 2PIGG sample.  
Furthermore, there is an objective prescription which relates the mass of a 
galaxy group in the 2PIGG catalogue to the mass of a dark matter halo in 
an N-body simulation (Eke et~al. 2004b). Careful tests using simulations show 
that there is a scatter and a small systematic bias between the true mass of 
the halo in the N-body simulation $m_{\rm true}$ and the estimated mass 
inferred from the galaxy groups found using a percolation algorithm 
(Eke et~al. 2004b):  
 
\begin{equation}
m_{\rm estimated} = m_{\rm true} \times 10^{ 0.1 + 0.3 \sigma}, 
\label{eq:merr}
\end{equation}
where the systematic bias is $0.1$\,dex and $\sigma$ is a Gaussian 
deviate with zero mean and unit variance. 
Hence, given the true halo mass obtained from the simulation, we can 
generate an estimated mass using Eq.~\ref{eq:merr}, to mimic the mass 
that would have been recovered for this halo by the 2PIGG algorithm.  

\begin{figure} 
       \includegraphics[width=80mm,height=80mm]{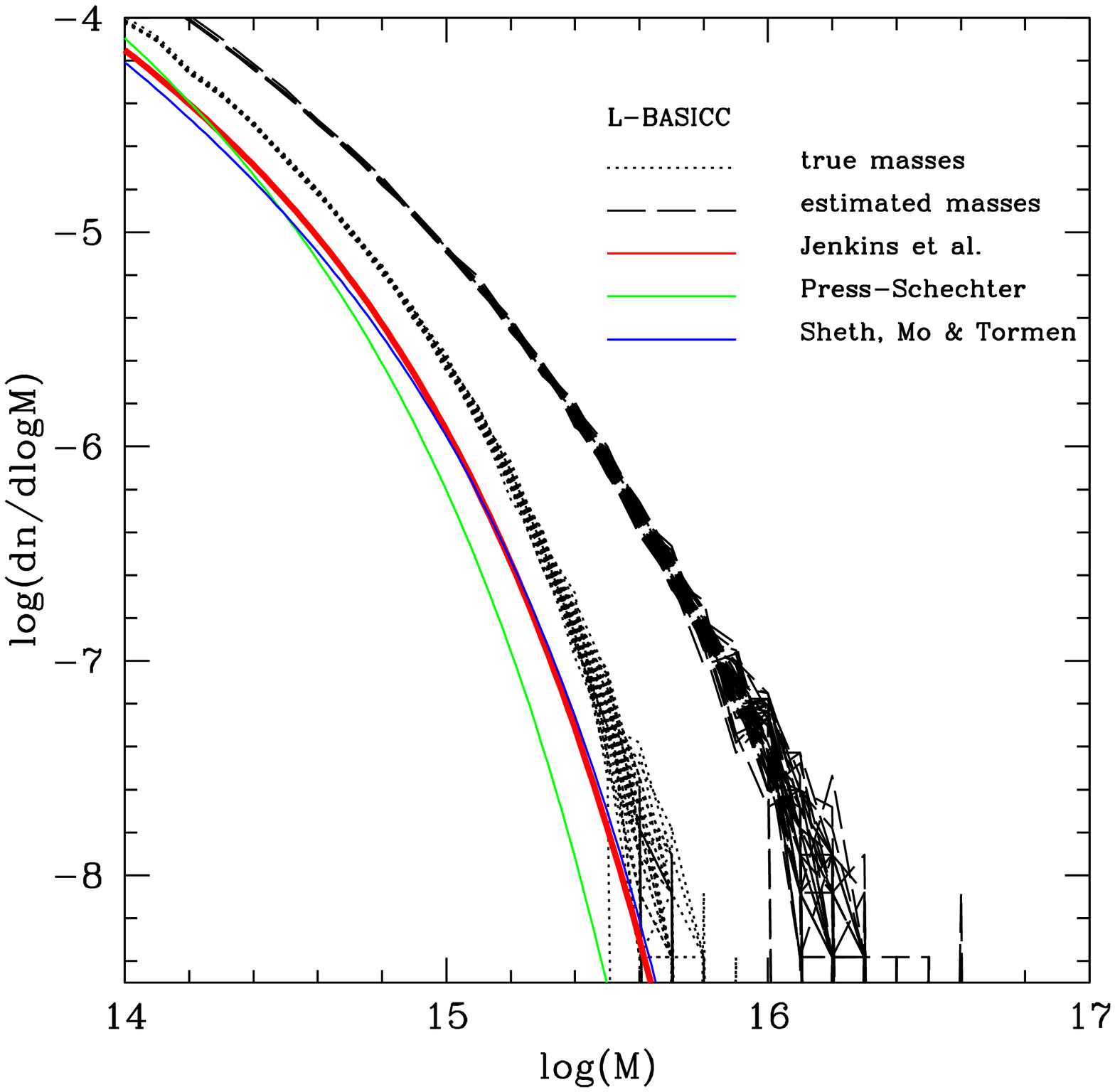}
       \includegraphics[width=80mm,height=80mm]{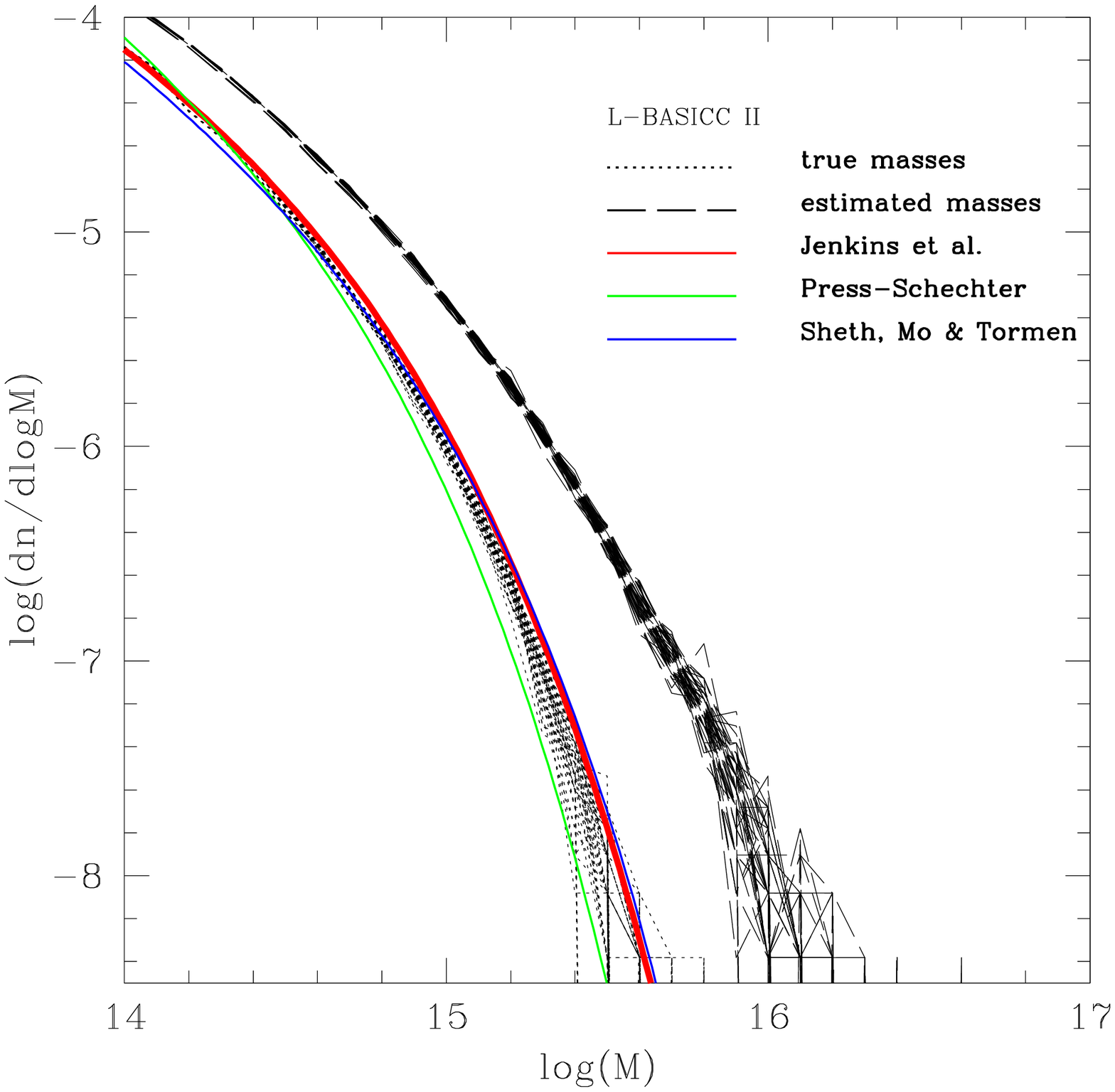}
\caption{ 
The abundance of dark matter haloes as a function of mass in the {\tt L-BASICC} (top) and 
{\tt L-BASICC II} (bottom) simulation ensembles. In both cases, the dotted black 
lines show the mass function using the true halo mass, as determined by the 
friends-of-friends group finder. The black 
dashed lines show the mass function after applying the prescription to generate the 
``estimated'' mass, as would be returned by applying the {\tt 2PIGG} algorithm (Eq.~2). 
Each line corresponds to a simulation from the ensemble. The coloured lines are the {\it same} 
in both panels and correspond to theoretical predictions for the true mass function in the 
{\tt L-BASICC II} cosmology, with the red curve showing the Jenkins et~al. (2001) 
empirical fit, the green curve the Press \& Schechter (1974) mass function and the 
blue curve the prediction of Sheth, Mo \& Tormen (2001). In the lower panel, these curves 
show how well the theoretical predictions agree with the simulation results; in the upper  
panel the same theoretical curves are plotted to show how much the simulation results 
have shifted following the change in cosmological parameters.}
\label{fig:mf}
\end{figure}

\begin{figure*} 
       \includegraphics[width=80mm,height=80mm,bb=80 30 430 370]{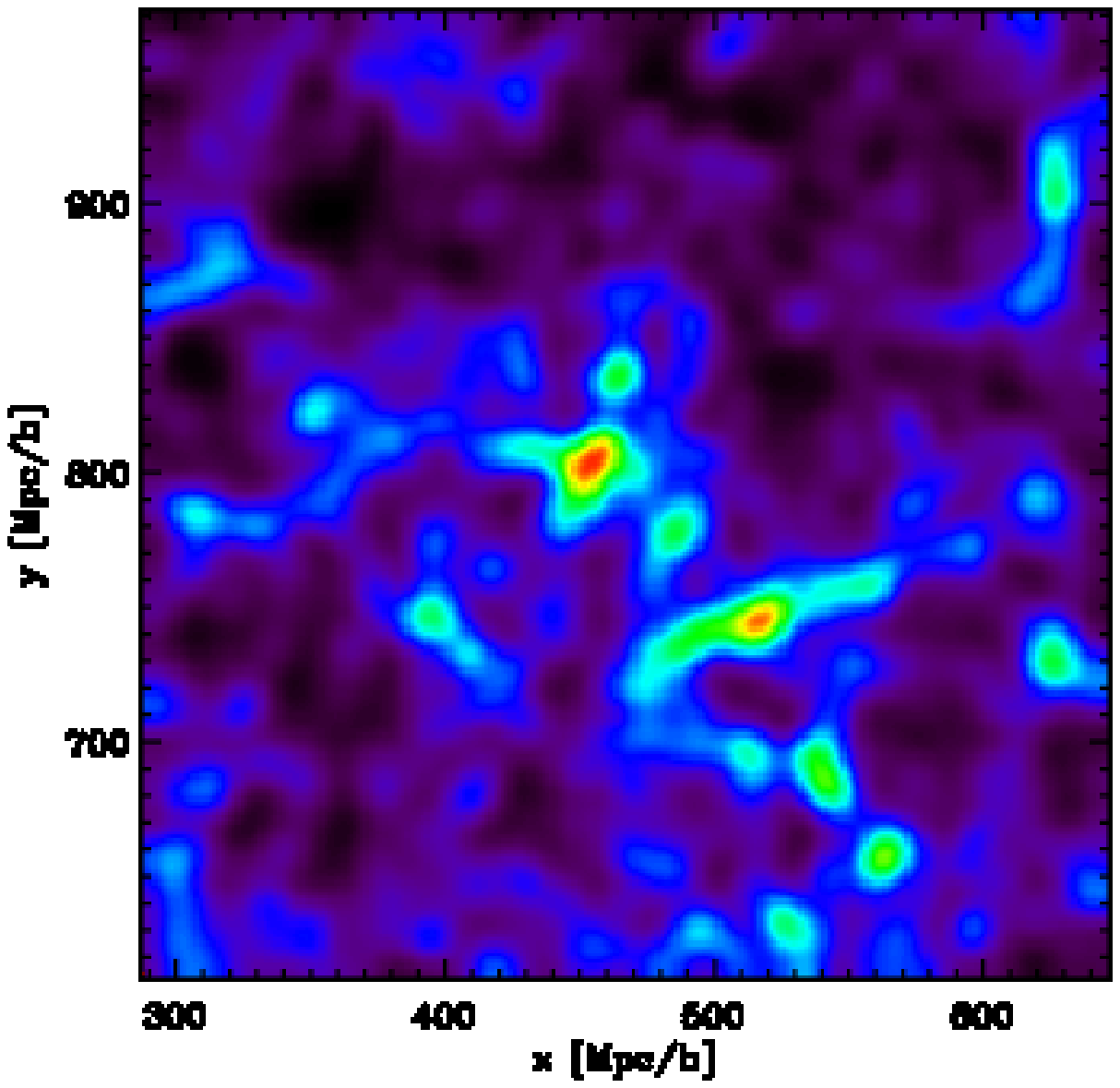}
       \includegraphics[width=80mm,height=80mm,bb=80 30 430 370]{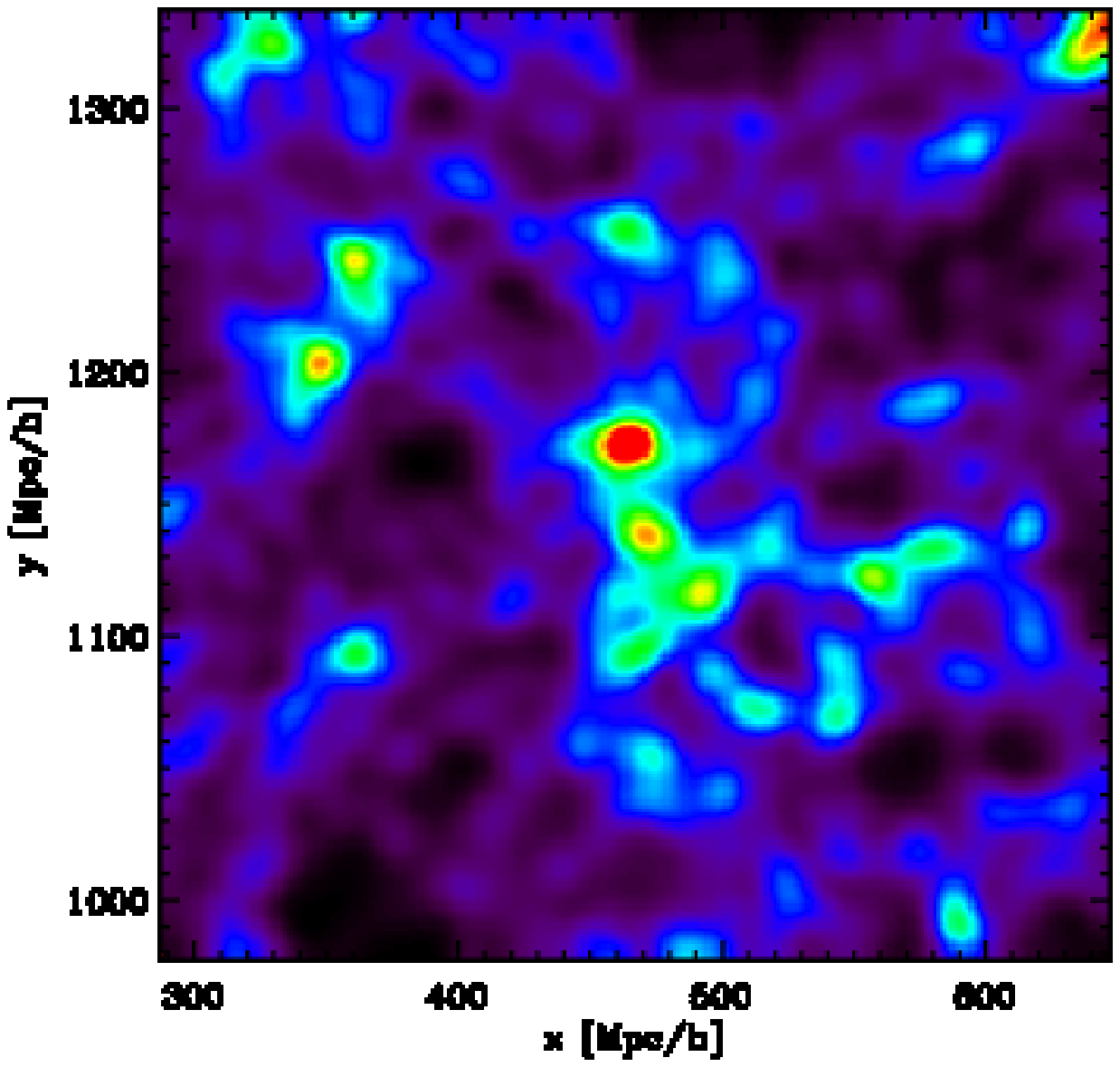}
       \includegraphics[width=80mm,height=80mm,bb=80 30 430 370]{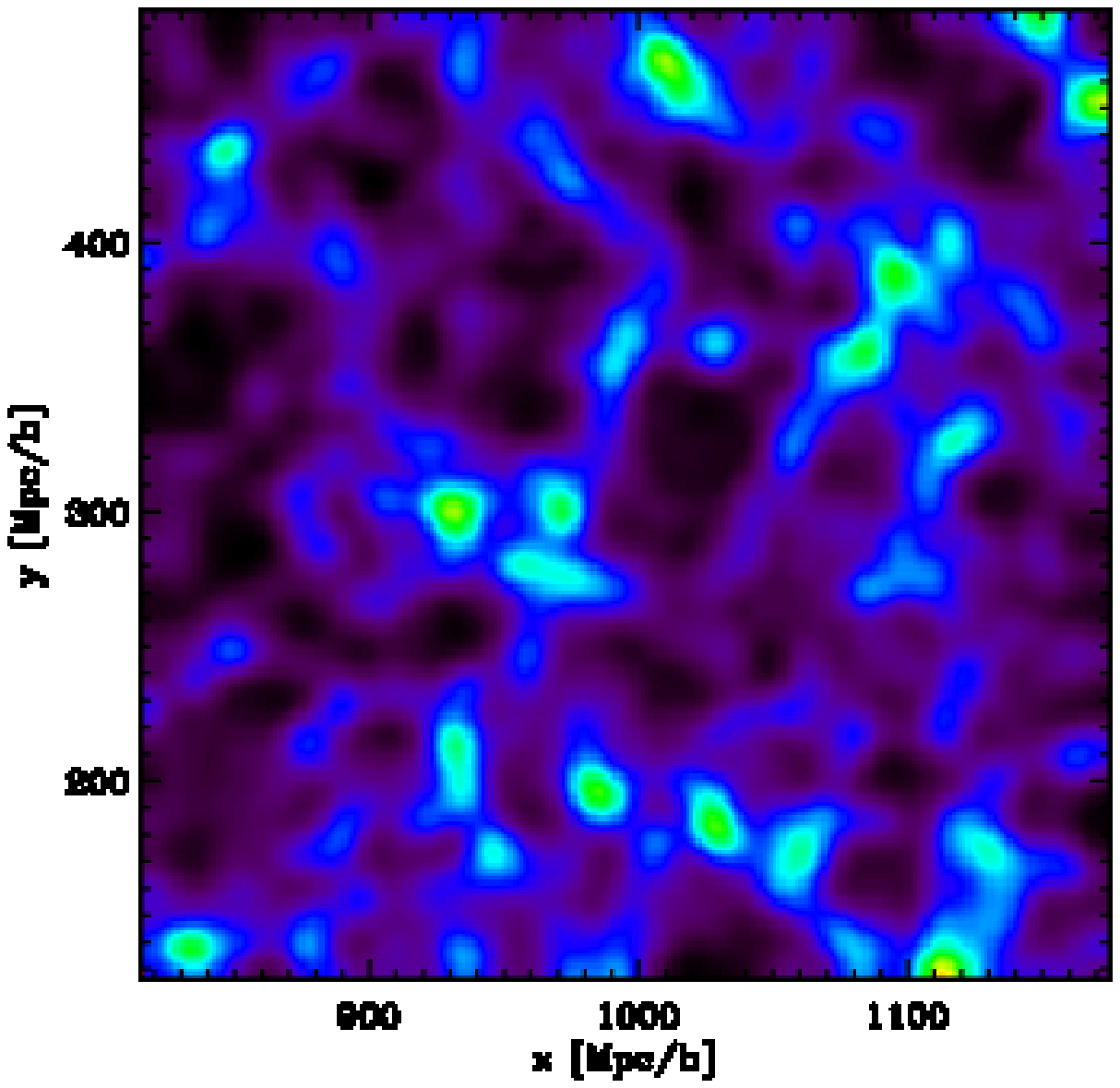}
       \includegraphics[width=80mm,height=80mm,bb=80 30 430 370]{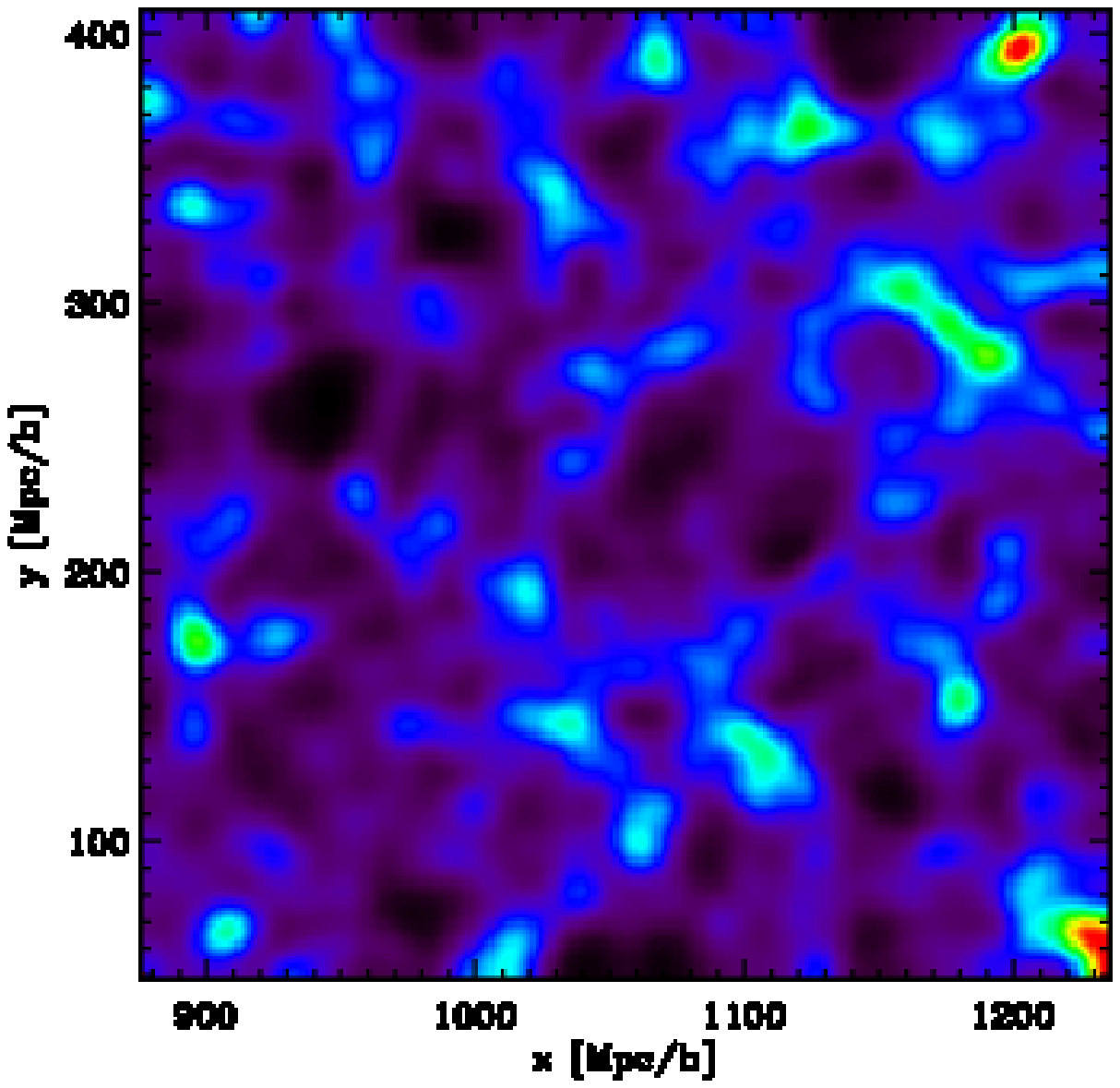}
\caption{
The projected density of dark matter in selected simulations. The colour scale
is the same in each panel and covers the range from 1 (black) to 16 (red) dark 
matter particles per pixel, with 256x256 pixels per image. Red corresponds to 
a projected dark matter density of  $1.5 \times 10^{13} h^{-1}M_{\odot}/(h^{-1} {\rm Mpc})^{2}$. 
The thickness of the slices is $40 h^{-1}$ Mpc. The top two panels are centred on 
the location of ``hot'' cells in the distribution of massive haloes, and the lower 
panels show regions chosen at random.
}
\label{fig:dens}
\end{figure*}

\subsection{The N-body simulations} 

We use two ensembles of 50 large volume N-body simulations 
called {\tt L-BASICC} (Angulo et~al. 2008). The ensembles 
correspond to different choices of the values of the cosmological 
parameters within a flat $\Lambda$CDM cosmology (Sanchez et~al. 2006, 
2009). The {\tt L-BASICC}\ ensemble uses the same cosmological 
parameters as the Millennium Simulation of Springel et~al. (2005): a 
matter density parameter, $\Omega_{\rm m}=0.25$, an energy density 
parameter for the cosmological constant, $\Omega_{\rm {\Lambda}}=0.75$, 
a normalization of density fluctuations, $\sigma_{8} = 0.9$, a 
Hubble constant of $h=0.73$, a scalar spectral index $n_{\rm s}=1$ 
and a baryon density parameter of 
$\Omega_{\rm b}=0.045$. The {\tt L-BASICC II} ensemble uses a 
parameter set which is in somewhat better agreement with the latest 
observations of the cosmic microwave background and the local large scale 
structure of the galaxy distribution (Sanchez et~al. 2006): 
$\Omega_{\rm m}=0.237$, $\Omega_{\rm b}=0.041$, $n_{\rm s} = 0.954$, $\sigma_{8} = 0.77$, and $h=0.735$.
Each of the {\tt L-BASICC}\ and  {\tt L-BASICC}\,{\tt II} simulations covers a comoving 
cubical region of side $1340\,h^{-1}\,$Mpc using $448^3$ particles. 
This gives a particle mass comparable to that employed in the Hubble 
Volume simulation (Evrard et~al. 2002). The equivalent Plummer softening 
length in the gravitational force is $\epsilon = 200\,h^{-1}\,\rm{kpc}$. 
The volume of each computational box, $2.41\,h^{-3}\,{\rm Gpc}^{3}$, is 
almost twenty times that of the Millennium Simulation, and more than three 
times the volume of the luminous red galaxy sample from the SDSS used to 
make the first detection of the acoustic peak by Eisenstein et~al. (2005). 
The computational box is 300 times the volume of the region covered by the 
volume limited 2dFGRS sample of $L_*$ galaxies. 
The total volume of the ensemble is $120\,h^{-3}\,{\rm Gpc}^{3}$, more than 
four times that of the Hubble Volume simulation. We use the friends-of-friends 
halo catalogue extracted from the $z=0$ simulation output in our analysis, 
retaining objects with ten or more particles (corresponding to a mass limit 
of $1.75 \times 10^{13}\,h^{-1}\,\rm{M_{\sun}}$) to analyze the likehood of 
finding rare massive structures.

\section{Results}

\begin{figure*} 
       \includegraphics[width=80mm,height=80mm]{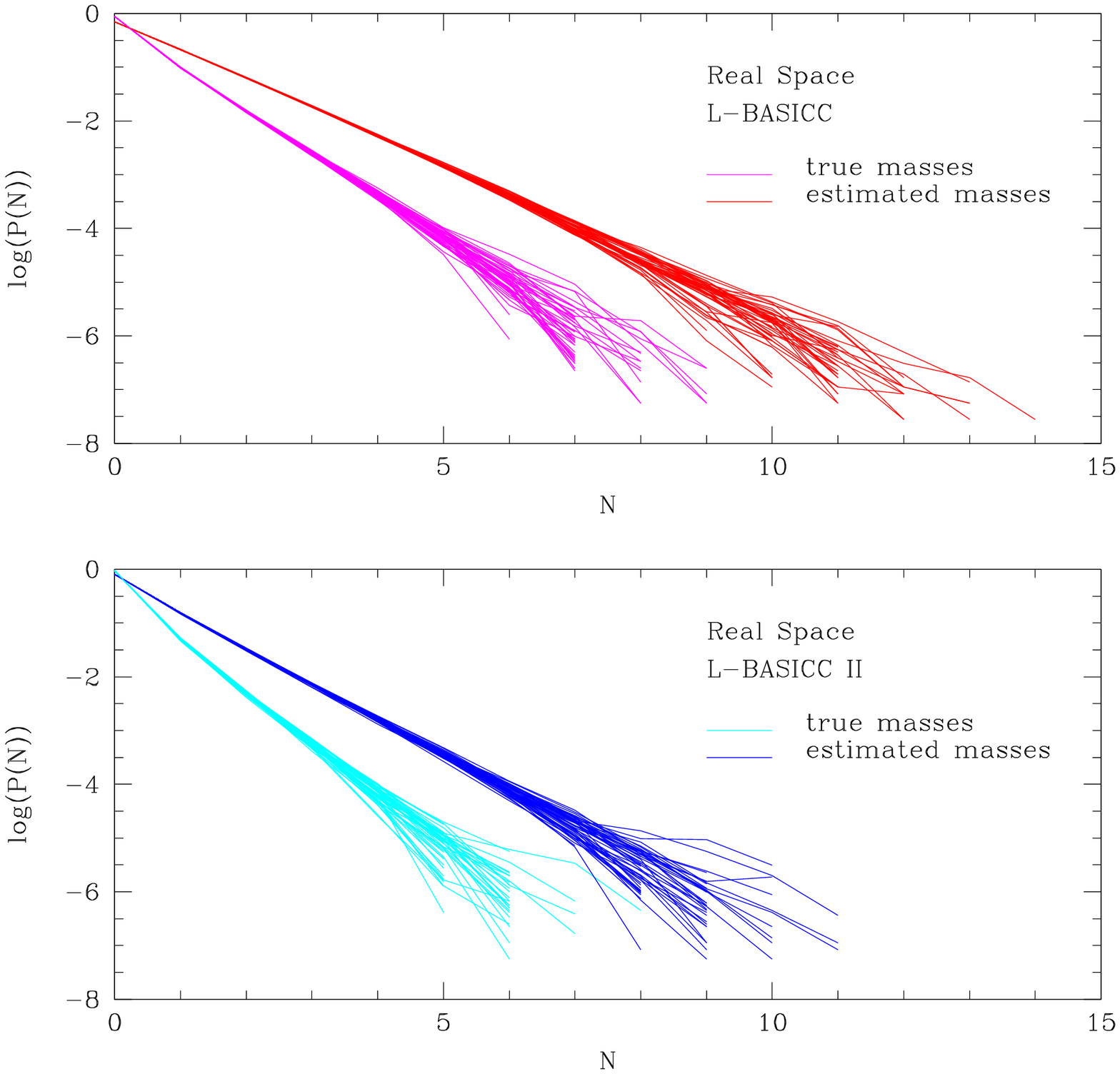}
       \includegraphics[width=80mm,height=80mm]{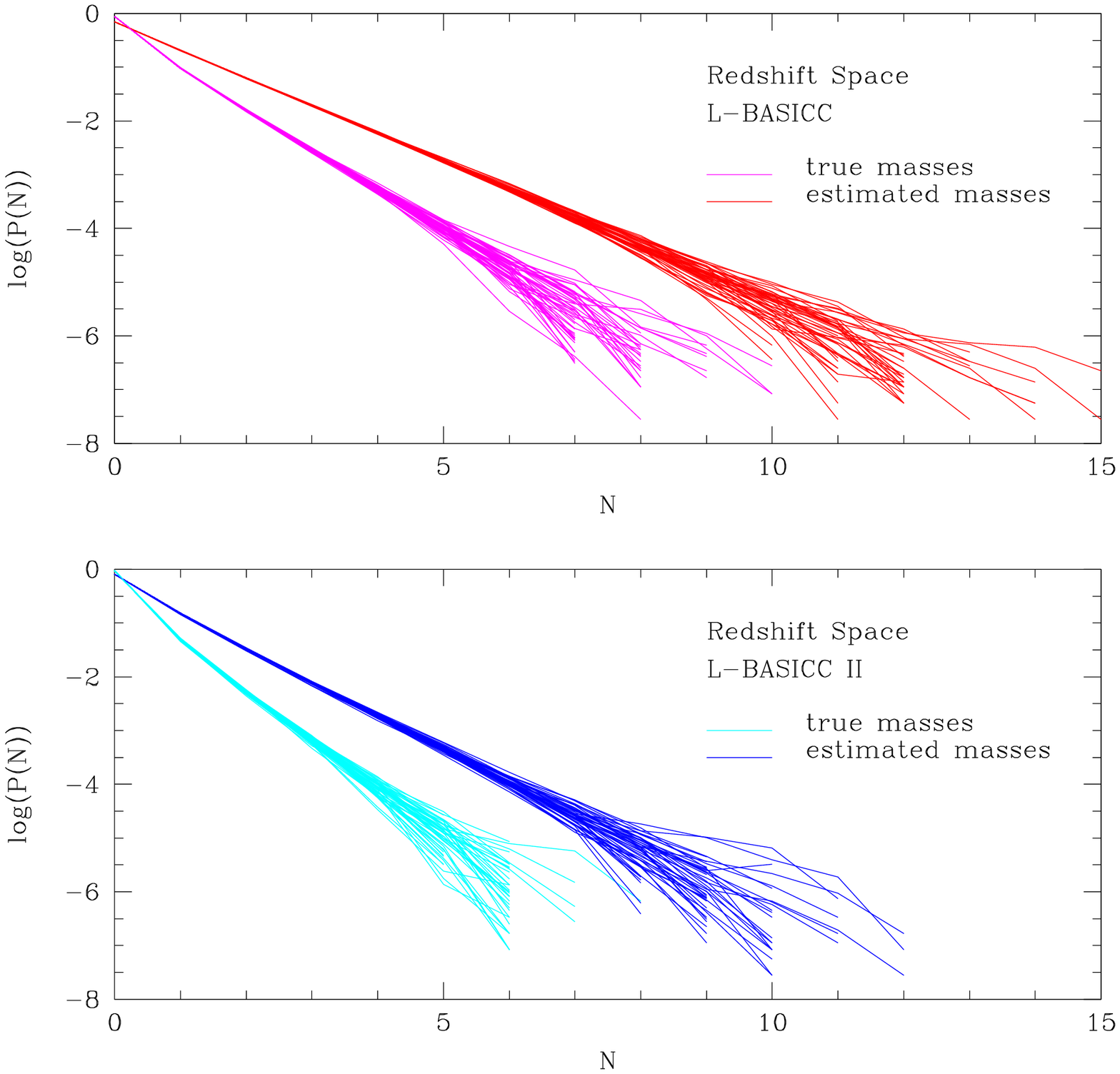}
\caption{
The count probability distribution for haloes with mass in excess of 
$5 \times 10^{14} h^{-1} M_{\odot}$ in cubical cells of side $40.6\,h^{-1}\,$Mpc.
In the left hand panels, the cell counts are measured using the real-space positions 
of the haloes and in the right-hand panels using the redshift-space positions. 
The upper row shows the measurement in the {\tt L-BASICC} ensemble and the 
lower row in the {\tt L-BASICC-II} simulations. Each curve shows the 
count distribution in one of the realizations. The different coloured lines in 
each panel show the measurements without and with mass errors as indicated 
by the key. 
}
\label{fig:pn}
\end{figure*}

\begin{figure} 
       \includegraphics[width=90mm,height=90mm]{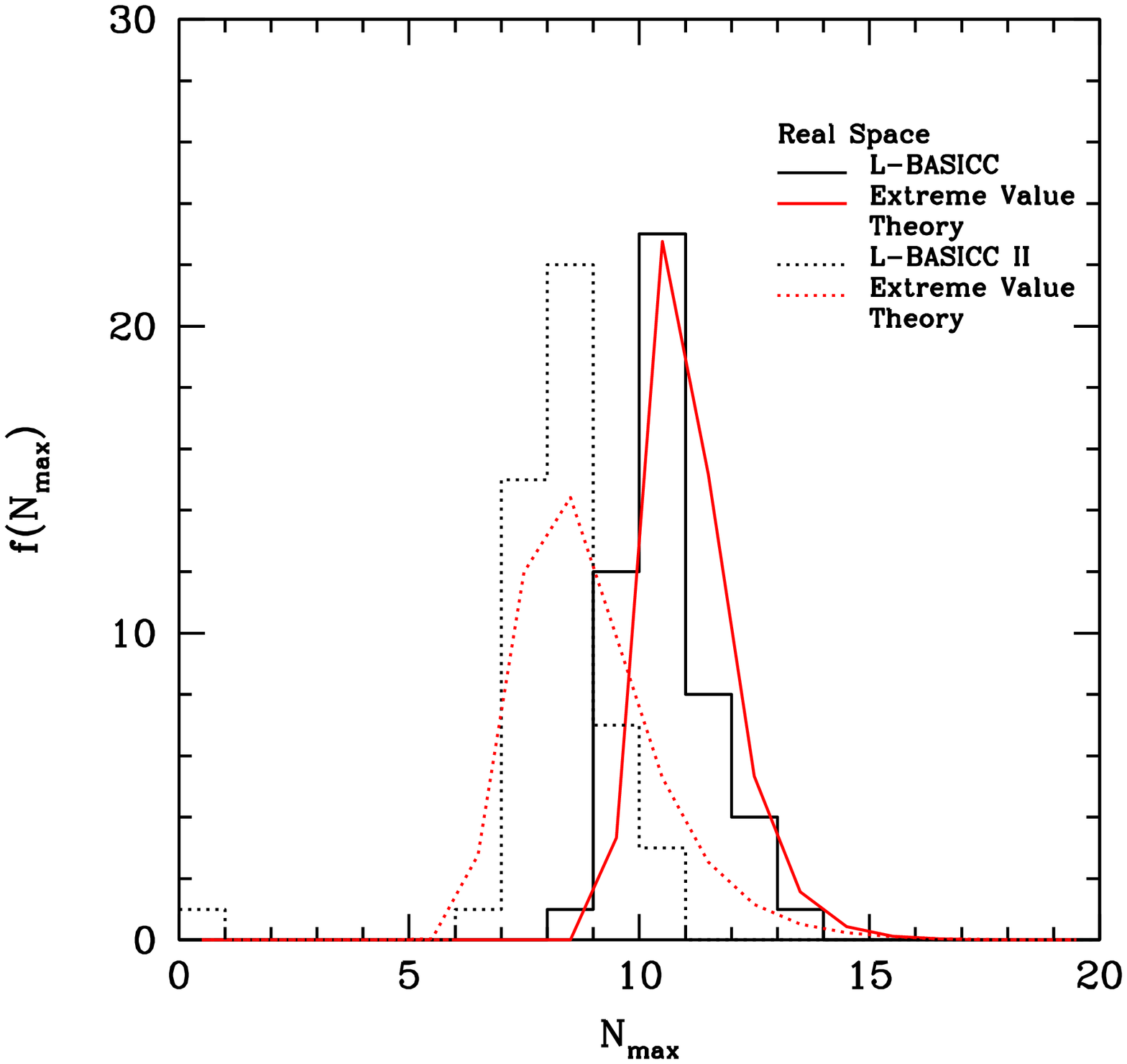}
       \includegraphics[width=90mm,height=90mm]{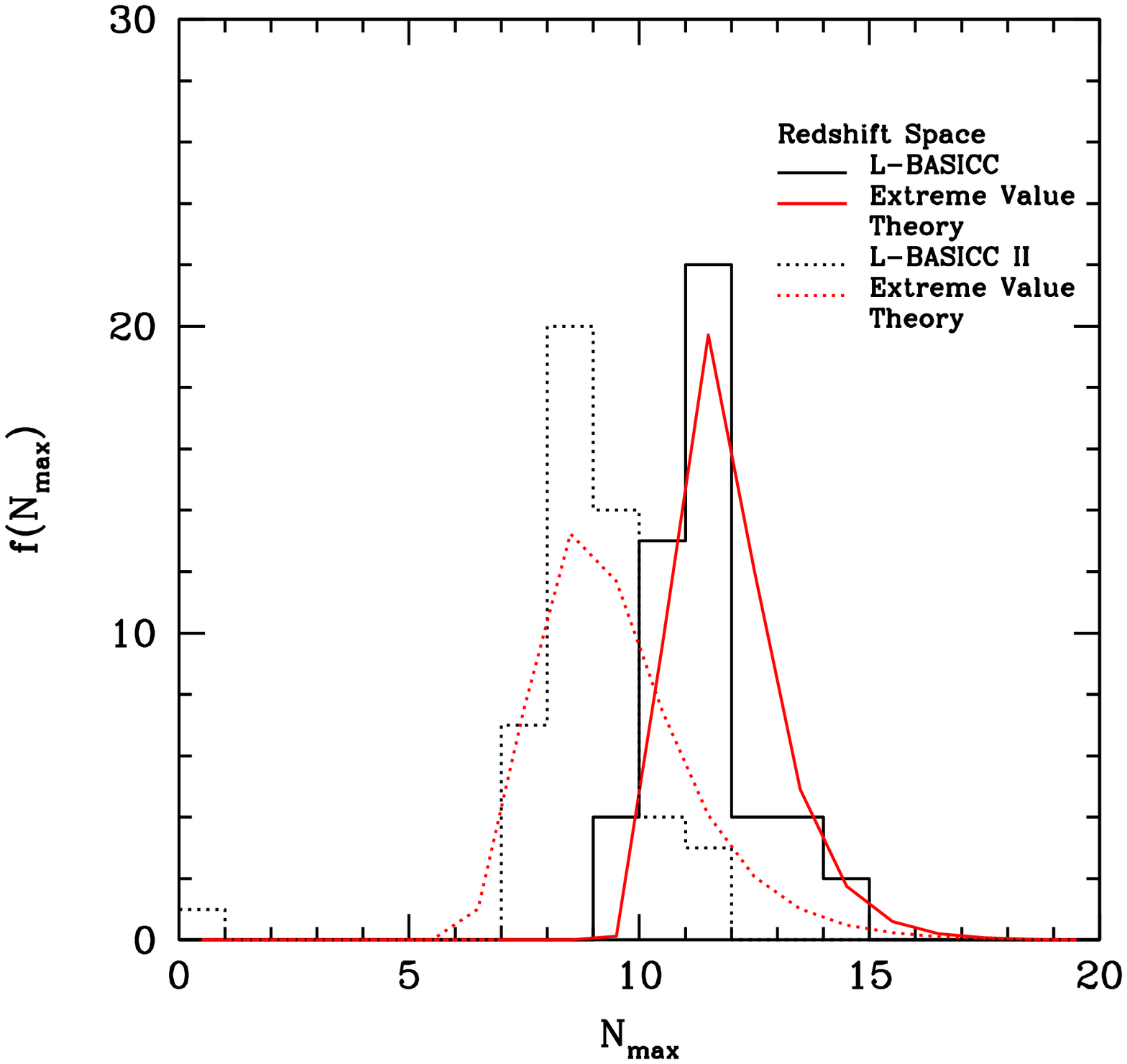}

\caption{
The distribution of the cell count in the hottest cells from each 
realization in the simulation ensembles. The number plotted on the 
x-axis is the number of haloes with mass $> 5 \times 10^{14}\, 
h^{-1}\,M_{\odot}$ in the cell. The upper panels show the cell counts 
measured in real-space and the lower panel shows the 
measurement in redshift-space. The results for the {\tt L-BASICC} 
ensemble are shown using solid histograms and for the {\tt L-BASICC-II} 
ensemble using dotted histograms. The corresponding Gumbel distributions 
are plotted in red. 
}
\label{fig:gumbel}
\end{figure}

In this section we present the results for the distribution of the counts in 
cells of dark matter haloes in the {\tt L-BASICC} and {\tt L-BASICC II} simulation 
ensembles. 

We first examine the impact on the halo mass function of including the bias and 
error expected if the {\tt 2PIGG} group finding algorithm was to be applied to 
a mock galaxy catalogue made from the N-body simulations. Fig.~\ref{fig:mf} shows 
the abundance of haloes at $z=0$, both without and with incorporating the error implied 
by the distribution given by Eq.~\ref{eq:merr}. There is a considerable spread in the 
halo mass function between the individual realizations of the ensembles for the most 
massive haloes. There is a significant change in the abundance of haloes at the high 
mass end of the mass function upon including the mass error. The abundance of haloes of 
mass $\log_{10}(M_{\rm halo}/h^{-1}M_{\odot}) \sim 15.5$ increases by an order of magnitude 
when the mass errors are included. This is readily understood in terms of the exponential 
shape of the mass function for haloes of these masses. There are many more haloes of 
low mass than high mass. Hence, on applying a perturbation in halo mass corresponding 
to a symmetrical error distribution, there is a net transfer of haloes from lower mass 
bins to higher mass bins. This effect is compounded by the small systematic overestimate 
of halo mass resulting from the {\tt 2PIGG} algorithm. The analytic mass functions give a 
reasonable match to the unperturbed mass function, except in the case of the Press \& 
Schechter theory, which predicts too few massive haloes. This discrepancy has been 
noted before (see e.g. Efstathiou et al. 1988). Finally, we note that there is an 
appreciable reduction in the abundance of haloes of a given mass in the {\tt L-BASICC II} 
cosmology compared with the {\tt L-BASICC} cosmology, as is clear by comparing the simulation 
results with the analytical models in Fig.~\ref{fig:mf}, which are the same in each panel.   

We now measure the distribution of the counts in cells of dark matter haloes in the 
simulations. We consider haloes with mass in excess of $5 \times 10^{14} h^{-1} M_{\odot}$, 
using both the true mass estimated directly from the simulation by a friends-of-friends halo finder 
using the dark matter particles and the perturbed mass, which attempts to mimic the mass the 
halo would have been assigned by the {\tt 2PIGG} algorithm. We use cubical cells of side $40.6\,h^{-1}\,$Mpc. 
To find cells of a similar halo overdensity to those corresponding to the 2dFGRS 
superclusters, we need to find cells which contain 10 or more haloes of the above mass. 
The halo distribution is oversampled by throwing down many more cells than would fit 
independently into the simulation volume. This is important because the halo count within 
a cell could change significantly with a small change in the location of the cell. This 
oversampling is taken into account when plotting the counts-in-cells probability distribution. 
We oversample the density field 1000 times; in practice the count probability 
distribution converges after a few tens of regriddings.

We show some images of selected regions in the {\tt L-BASICC} simulations in Fig.~\ref{fig:dens}. 
These plots show the projected density of dark matter in slices of $40 h^{-1}$Mpc. The volume 
in these slices is 50\% of the volume of the 2dFGRS $L_*$ sample. The top panels 
are centred on ``hot'' cells which contain 10 or more haloes more massive than 
$5 \times 10^{14} h^{-1} M_{\odot}$ and the lower panels show randomly selected regions.

We first measure the cell counts using the true positions of the haloes, without taking into 
account the impact of peculiar velocities. The cell counts for the two ensembles of simulations 
are plotted in Fig.~\ref{fig:pn}. In all 100 realizations covering both cosmologies, we did 
not find any cells with the required occupancy of massive haloes to match the 2dFGRS superclusters, 
when using the true halo masses. The conclusion is very different if we consider the estimated halo 
masses rather than the true halo masses. Now there are many cells with 10 or more haloes of the 
required mass. In the {\tt L-BASICC} cosmology, in the majority of realizations the hottest cells 
contain in excess of 10 massive haloes. Around a quarter of the realizations in the {\tt L-BASICC-II} 
ensemble contain cells with counts above this threshold. This difference in the tail of the count 
distribution is driven primarily by the difference in the value of $\sigma_{8}$ between {\tt L-BASICC} 
and {\tt L-BASICC II}. 

The impact of peculiar velocities on the appearance of large scale structure in the halo distribution 
is taken into account using the distant observer approximation. One of the cartesian axes of the 
simulation cube is taken to be the line of sight. The peculiar motion of the halo along this 
axis is added to its position, after applying a suitable scaling to convert the velocity into an 
equivalent displacement in Mpc. The change in the cell count probability is quite dramatic, as can be 
seen from Fig.~\ref{fig:pn}. There is a shift in the maximum cell count and an increase in the 
spread of the maximum cell count across the realizations of the simulation ensembles. The impact of 
redshift space distortions on the cell counts is akin to changing 
cosmology from that used in the {\tt L-BASICC} to {\tt L-BASICC-II} 
simulations (which is largely due to the change in the value of $\sigma_8$). 

In order to quantify these changes more clearly, we plot in Fig.~\ref{fig:gumbel} the distribution 
of the maximum cell count over the different realizations. The histograms show the distribution of hot cell 
counts extracted from the simulations. The curves show the Gumbel distribution of Eq.~1, 
plotted using the corresponding mean and variance of the hot cell distributions as 
measured from the simulations. In real-space, the mean and variance of the hot cells 
for the {\tt L-BASICC} ensemble is 11.1 and 0.995 respectively, which correspond to 
$\mu = 10.65$ and $\beta=0.78$; in the case of {\tt L-BASICC-II} these values become 
a mean of 8.92 and variance of 1.562, which become $\mu = 8.21$ and $\beta=1.22$.
The agreement between the Gumbel distributions and the simulation results is impressive, 
being equally good in real and redshift space and with and without mass errors.

\begin{figure} 
       \includegraphics[width=80mm,height=80mm]{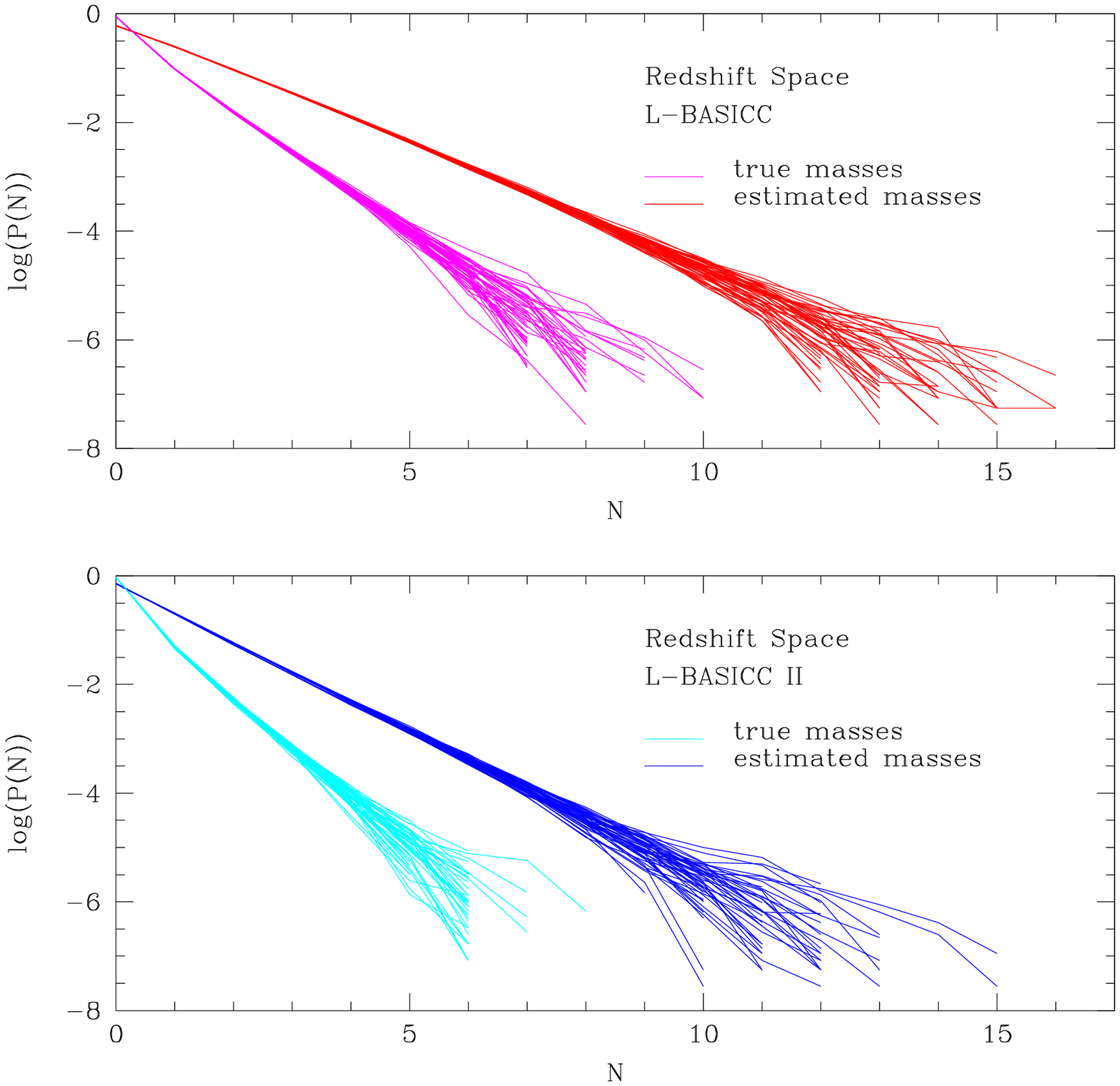}
       \includegraphics[width=80mm,height=80mm]{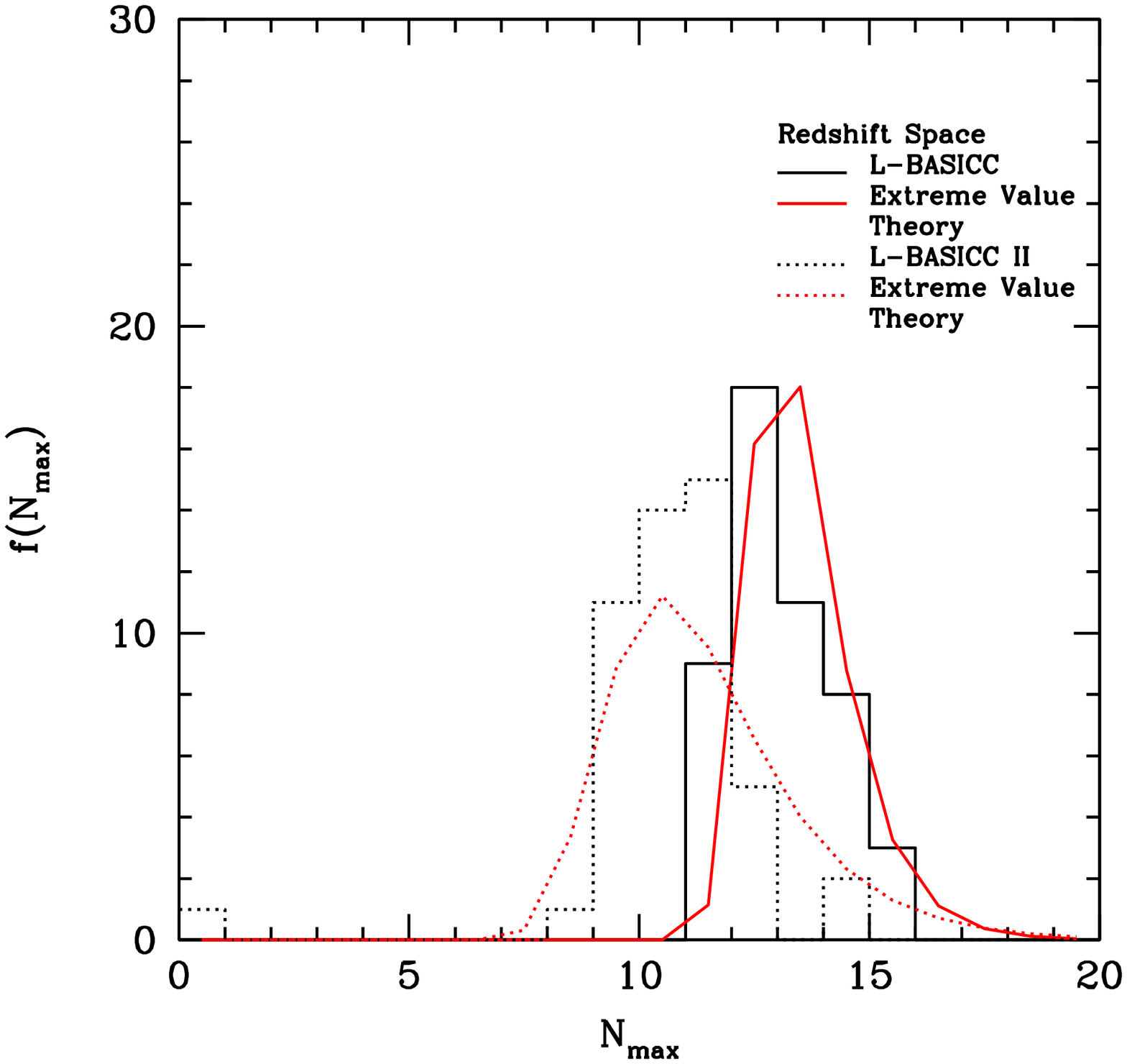}
\caption{  
The impact of increasing the size of the scatter in the recovered halo mass. The 
upper two panels show the cell counts in redshift space, using the true halo 
masses (magenta and cyan lines in the top and middle panels, respectively) and 
with increasing the scatter in the mass from the value of $0.3$~dex used elsewhere 
in this paper to $0.4$dex (red and blue curves). The upper and middle panels show 
the results for different cosmologies as labelled. The resulting distributions of 
maximum cell counts are shown in the bottom panel.
}
\label{fig:err}
\end{figure}

Our results are sensitive to the assumed error on the halo mass returned by the 
group finding algorithm applied to the 2dFGRS. In Fig.~\ref{fig:err} we show the impact on 
the cell count probability of a modest increase in the size of the scatter in halo 
mass, using $0.4$~dex instead of $0.3$~dex. This 25\% increase in the mass error produces a 
substantial shift in the tail of the count distribution and, correspondingly, on the 
distribution of maximum cell counts, shown in the lower panel of Fig.~\ref{fig:err}. Eke et~al. (2004b) 
made a careful assessment of the error on the recovered mass of 2dFGRS groups, using N-body 
simulations combined with a galaxy formation model. However, it is not inconceivable that the 
parameters used in the group finding algorithm could have some dependence on the 
galaxy formation model used in the calibration, as could the error on the recovered masses, 
particularly at the $25\%$ level.

So far we have assessed the chance of finding a hot cell within the full simulation volume which is 
many times larger than the volume of the 2dFGRS $L_*$ volume limited sample. We have found that cells 
with as many massive haloes as found in the 2dFGRS are not uncommon in the current best fitting 
CDM models, particularly when the clustering is measured in redshift space and the effects of mass 
errors introduced by the group finding procedure are included. The volume of the $L_*$ 
sample is equivalent to around 120 of the cubical cells of side  $40.6\,h^{-1}\,$Mpc.
We have measured the probability of finding a hot cell in groups of 120 cells drawn from the simulations. 
The cells are contiguous in the $x$ and $y$ planes within the simulation cube, but do not represent an 
attempt to extract a region with the same geometry as the 2dFGRS itself, just the same volume. In the 
{\tt L-BASICC-II} ensemble in redshift space, we find that 
in around $2$\% of cases, a hot cell is found. Hence, the chance of finding a hot cell like the ones 
in the 2dFGRS is low but not negligible. 

\section{Conclusions}

We have introduced a new objective methodology for assessing the likelihood of finding extreme  
structures in hierarchical structure formation models. Quite often the probability of finding an 
unusual structure such as a large void or an overdensity is estimated using a Gaussian distribution, 
because the smoothing scale in question is large. This is a good approximation for events which 
represent small departures from the mean density. However, for extreme events this is a poor 
assumption. The probability distribution of the density contrast on a particular smoothing 
scale, although assumed to be initially Gaussian in most models, rapidly evolves away from 
this form due to gravitational instability (e.g. Gazta\~{n}aga, Fosalba \& Elizalde 2000). 
Assuming a Gaussian probability distribution rather than the true distribution would lead to 
a misestimation of the probability of finding a cell with an extreme density by many orders 
of magnitude. 

The advantage of our approach is that we do not need to specify the actual form of the 
probability distribution of cell counts. We have shown that the distribution  of extreme 
cell counts is well described by a Gumbel distribution in a range of different 
situations: real-space, redshift-space and both with and without errors in halo mass. 
The simulations give the mean and variance of the Gumbel distribution. The analytic form 
can then be extrapolated into the tails to give the probability of events which would require 
hundreds or even thousands of realizations of N-body simulations to find. By using N-body simulations, 
we can assess the probability of events which cannot be calculated analytically, such as the 
largest Einstein ring expected in a CDM model (e.g. Oguri \& Blandford 2009). 

Cells with the number of massive haloes seen in the 2dFGRS can be found 
within our simulations, provided that the clustering of these haloes is 
measured in redshift-space and the mass errors introduced by the group 
finding algorithm are taken into account. However, if we consider a volume 
of the size of the 2dFGRS $L_*$ sample, which is 300 times smaller than our 
simulation volume, then we expect to find such an overdensity of cluster mass 
haloes in $\approx 2$ out of a hundred cases. 

Norberg et al. (2010) have carried 
out a related analysis using the final release of the SDSS. These authors applied 
a different technique to identify overdense regions. They split the galaxy distribution 
into zones, as would be done to carry out a jackknife estimation of the error on 
clustering statistics (Norberg et~al. 2009). By comparing the distribution of two and three point 
correlation functions measured from the jackknife resamplings, a zone whose omission produced 
an outlying estimate of the clustering was found. However, when applying the same 
analysis to the same ensemble of N-body simulations used in this paper, Norberg et~al. 
(2010) found that such outliers were quite common. There are some key differences 
between our analyses. The SDSS volume limited sample is an order of magnitude larger 
than the 2dFGRS sample considered in this paper. Norberg et al. found one ``unusual'' structure 
in the SDSS volume limited sample for $L_*$ galaxies. Also, the method for quantifying unusual structures 
is different from the one we use, and will pick up a very different type of structure. 
The zones used by Norberg et~al. sample conical volumes of space, covering a large 
baseline in radial distance. The superstructure in their case could be a projection of 
independent structures along the line of sight. In our case, we use compact cells. 
Norberg et~al. (2010) conclude that in the larger volume of the SDSS $L_*$ sample, 
the structures found by their correlation function study are consistent with those 
found in CDM. Our results do not contradict this; we have applied a different 
test to search for overdense regions with a different structure.

\section*{Acknowledgements}
YY acknowledges a studentship from CONICET. This work was supported by the European 
Commission through its funding of the LENAC training project under the ALFA-II program 
and by the Science and Technology Facilities Council. We acknowledge Stephane Colombi, 
Peter Coles and John Barrow for stimulating discussions on extreme value theory and the 
referee for a helpful report.

\end{document}